\documentstyle[aps,prl,multicol]{revtex}
\begin{document} 
\input{psfig.tex}
\draft

\title{Pseudogap phase in the U(1) gauge theory with incoherent spinon pairs}
\author{Xi Dai, Yue Yu, Tao Xiang and Zhao-bin Su}
\address{Institute of Theoretical Physics, Chinese Academy of Sciences, Beijing
       100080, P.R.China}
\date{\today}       
\maketitle 

\begin{abstract}

The pseudogap effect of underdoped high-$T_c$ superconductors is studied in
the U(1) gauge theory of the t-J model including the spinon pairing
fluctuation. The gauge fluctuation breaks the long range correlation between
the spinon pairs. The pairing fluctuation, however, suppresses significantly
the low-lying gauge fluctuations and leads to a stable but phase incoherent
spin gap phase which is responsible for the pseudogap effects. This quantum
disordered spin gap phase emerges below a characteristic temperature $T^{*}$
which is determined by the effective potential for the spinon pairing gap
amplitude. The resistivity is suppressed by the phase fluctuation below $%
T^{*}$, consistent with experiments.

\end{abstract}
\vskip 0.1in
 
%\hspace{1cm}

\begin{multicols}{2}

The normal state behavior of high-$T_c$ cuprates has cast doubt on the
conjecture that correlated electron metals are Fermi liquids in the absence
of symmetry breaking. One anomalous feature which is difficult to understand
is the spin gap or pseudogap behavior\cite{batlogg}
observed in underdoped materials. This
phenomenon was first observed in the NMR measurement\cite{nmr1,nmr2} 
of deoxygenated YBCO,
which showed that both the spin susceptibility and the nuclear spin-lattice
relaxation rate were suppressed below some temperature above $T_c$,
indicating an opening of a spin excitation gap. This anomalous behavior of
underdoped materials also manifests itself in the resistivity\cite{res1,res2}, 
Hall coefficient\cite{hall}, 
specific heat\cite{spec1,spec2}, optical properties\cite{opt},
inelastic neutron scattering\cite{iens}, 
and other thermodynamic or transport quantities\cite{tun1,tun2}.

At present there is no consensus as to the correct theory of the pseudogap
effect. One interpretation is that the pseudogap is the energy gap of
pre-formed Cooper pairs\cite{ek,ran}, 
and phase fluctuations in the underdoped regime
prevent these pairs condensing until it has reached some lower temperature.
This type of theory gives a qualitative account for the similarity between
the pseudogap and the superconducting gap. But it is not clear if the phase
fluctuation is really strong enough to surpress $T_c$ by one order of
magnitude in the extremely underdoped limit.

A different interpretation is given by the resonant valence bond mean
field theory\cite{UL1,MF} 
based on the notion of charge-spin separation: electrons
separate into spinons and holons, spinons have spin but no charge, holons
have charge but no spin. In this theory, the spinons are predicted to form
singlet pairs well above the superconducting $T_c$, with superconductivity
setting in only when the holons become phase coherent at $T_c$. This theory
provides a qualitative description of the high-$T_c$ phase 
diagram\cite{UL1,MF}. However,
the spin gap phase is unstable against the fluctuation of the U(1) gauge
field\cite{UL2}, 
which is introduced to enhance the constraints of no double occupancy.

In this paper, we study the effect of the spinon pairing fluctuation on the 
physical properties of the spin gap phase in the U(1) gauge theory of 
the t-J model. we propose
that instead of the mean field spinon pairing phase with long range phase
conference, a local pairing phase of spinons which has no long range
phase coherence will survive under the fluctuation of U(1) gauge field
because it keeps the local U(1) symmetry. 
Then such a description can be viewed as the strong coupling
version or the microscopic origin of the nodal liquid proposed by 
Balents,Fisher and Nayak(BFN)\cite{bfn}.
The reason which quantum disorders the d-wave pairing state is attributed
to the fluctuation of U(1) gauge field, which is absent in BFN's theory.
Since the U(1) gauge field is added to insure the local constrain in
t-J model, the quantum disordering of the spinon pairing state is also
a result of the local constrain. Firstly, the spinon pairing state should
be regarded as the mean field description of the RVB state. In such a
description the phase of the pairing order parameter is fixed, because
the phase operator is conjugate to the particle number 
operator, the fluctuation
of the occupy number will be infinitive due to the uncertainty principle.
Since actually this kind of fluctuation is very small in Under doped regime
because of the local constrain, when we go beyond the mean field theory
by considering the fluctuation of the U(1) gauge field it is quite 
reasonable to have a result with large uncertainty in the phase and
small uncertainty in the occupy number. Then the instability problem 
of the spinon pairing phase is physically caused by the local constrain
and should be cured by disordering the phase of the order parameter.

After integrating out the spinon and holon operators, we obtain an
effective action for the gauge and pairing fields ${\cal S}={\cal S}_\Delta +
{\cal S}_A$.
Then the total action is ${\cal S}={\cal S}_\phi +{\cal S}_A$.

The effective action for the gauge field\cite{UL2,LN} is given by
\begin{equation}
{\cal S}_A={\frac T2}\sum_{i\nu_n}\int {\frac{d^2q}{(2\pi )^2}}\Pi_{\mu
\nu }({\bf q},i\nu_n)A_q^\mu A_{-q}^\nu , 
\end{equation}
where $A^\mu $ is the U(1) gauge field. The inverse of the gauge field
propagator is\cite{LN}
\begin{equation}
\Pi_{\mu \nu }({\bf q},\nu)=\chi_dq^2-i\gamma_F\nu /q, 
\end{equation}
where $\gamma_F=2n_e/k_F$ and $\chi_d=\chi_F+\chi_B$. $\chi_F$ and $
\chi_B$ are the Landau diamagnetic susceptibilities of spinons and holons,
respectively.

In order to treat the fluctuation of pairing order parameter, we
assume that the dynamics of the spinon pairing order parameter is described by an
effective Ginsburg-Landau theory\cite{LN2} 
with a minimal coupling with the gauge
field, namely
\begin{equation}
{\cal S}_\Delta =\int dx\left[ \widetilde{\kappa }_\mu \Delta^{*}(x)\left(
\partial_\mu -A_\mu (x)\right)^2\Delta (x)+\alpha |\Delta |^2+\beta
|\Delta |^4\right] ,
\end{equation}
where $\widetilde{\kappa }_0={1\over{2m^{*}c^2}}$, 
$\widetilde{\kappa }_{1,2}={1\over{
2m^{*}}}$, $m^{*}$ and $c$ are the effective mass and velocity of spinon
pairs, respectively.  We further assume that 
$\alpha =a(T-T_0^{*})$ and $\beta $
is independent of temperature. In $\alpha$, $a$ is a constant and $T_0^{*}$ is the
critical temperature of the spin gap phase without gauge fluctuation. 

In the work of Ubben and Lee\cite{UL2}, 
the phase fluctuation is ignored. The energy
loss due to the gauge fluctuation is proportional to $\Delta^{5/3}$, while
the energy gain from spinon pairing is proportional to $\Delta^2$. Thus for
small $\Delta $ the first term always dominates and the phase coherent spin
gap phase is unstable. This means that to study the pseudogap phase in
the U(1) gauge model, the pairing phase fluctuation should be considered. 

The spinon pairing field $\Delta (x)=\sqrt{\rho }e^{-i\varphi }$ 
contains both the
amplitude and the phase fluctuations. When $T\ll $ $T_0^{*}$, the amplitude
fluctuation is massive and can therefore be omitted. However, when $T\simeq
T_0^{*}$, the amplitude fluctuation is important. If we assume that $\rho
(x) $ varies slowly in both time and space, then ${\cal S}_\Delta $ is
approximately given by
\begin{equation}
{\cal S}_\Delta \approx \int d^3x\left[ {\frac 12}\kappa_\mu (x)\left(
\partial_\mu \varphi (x)-A_\mu (x)\right)^2+\alpha \rho (x)+\beta \rho
(x)^2\right] ,
\end{equation}
where $\kappa_\mu (x)=\rho (x)\widetilde{\kappa }_\mu $. To treat the 
phase fluctuation properly, we introduce the duality 
transformation, fllowing reference \cite{bfn}.
The phase variable 
$\varphi $ is generally multivalued and $\partial_\mu \varphi (x)$ is not
curl-free. Thus there are vortices in the $\varphi $ field and the cure of $
\partial_\mu \varphi (x)$ defines a vortex current operator  $j_\mu
^v=\epsilon_{\mu \nu \lambda }\partial_\nu \partial_\lambda \varphi $. To
treat these vortices, a commonly used approach is to introduce a fictitious
gauge field $a_\mu $, which is dual the phase variable $\varphi $, via the
equation  
\begin{equation}
\kappa_\mu (\partial_\mu \varphi -A_\mu )=\epsilon_{\mu \nu \lambda
}\partial_\nu a_\lambda .
\end{equation}
Substituting the solution of $\varphi $ from the above equation to the
definition of the vortex current, one can relate $a_\mu $ to the vortex
current operator 
\begin{equation}
j_\mu^v=\epsilon_{\mu \nu \lambda }\partial_\nu \left[ \kappa_\lambda
^{-1}\epsilon_{\lambda \alpha \beta }\partial_\alpha a_\beta +A_\lambda
\right] .
\end{equation}

In the dual representation of $\varphi $, vortices are treated as quantized
particles and represented by a complex field $\Phi $\cite{bfn}. 
A dual Lagrangian of $
\varphi $ with minimal coupling between $a_\mu $ and $\Phi $ can be
constructed as 
\begin{eqnarray}
{\cal L}_\varphi &=&\epsilon_{\mu \nu \lambda }{\frac 1{{2\kappa_\mu (x)}}}
(\partial_\nu a_\lambda -\partial_\lambda a_\mu )^2+\epsilon_{\mu \nu
\lambda }a_\mu \partial_\nu A_\lambda  \nonumber \\
&&+{\frac{\kappa_\mu (x)}2}|(\partial_\mu -ia_\mu )\Phi |^2-V_2\Phi^2,
\end{eqnarray}
where the higher order terms of $\Phi $ are ignored. From the equation of
motion of ${\cal L}_\varphi $, it can be shown that this Lagrangian has the
desired property of the vortex current defined above. The bare mass of the
vortex field $V_2$ depends on the superfluid density of spinon pairing $\rho
(x)$ and can be estimated as follows. Close to the transition temperature $
T_0^{*}$, the excitation energy of a single static vortex is approximately
equal to $E_V\approx \sqrt{2V_2/\kappa_0}$. From the GL theory, we know
that $E_V$ is also proportional to the spinon superfluid density $\rho (x)$
\cite{dg}
if we assume the spinon superconductivity is also type II,
i.e. $E_V=g\rho (x)$ and $g$ is a constant. Thus $V_2$ is approximately
equal to 
\begin{equation}
V_2\approx {\frac 12}g^2\rho^2\kappa_0={\frac{g^2}{{4m^{*}c^2}}}\rho (x)^3,
\label{V2}
\end{equation}

The highest power of $a_\mu $ in ${\cal L}_\varphi $ is 2. $a_\mu $ can
therefore be rigorously integrated out from the Lagrangian. This leads to a
self energy correction to the propagator of the gauge field $A_\mu $: 
\begin{equation}
\Pi_{\mu \nu }^{\prime }({\bf q},{\bf q^{\prime }},\nu )=\Pi_{\mu \nu }(
{\bf q},\nu )\delta_{{\bf q},{\bf q^{\prime }}}+{\frac{q_\mu q_\nu^{\prime
}}{{\frac 12}\sum_k\kappa_0(k)\Phi^2(-k+q^{\prime }-q)}.} 
\end{equation}

By further integrating out the gauge field ${\bf A}$, an effective actioin
for the $\Phi (x)$ and $\rho (x)$ fields is obtained as\cite{UL2}
\begin{eqnarray}
S\left[ \Phi (x),\rho (x)\right]  &\approx &-f_0\left[ 1+\frac 1{{\int d^2x{
\frac 12}\chi_d\kappa_0(x)\Phi^2}(x)}\right]^{-{\frac 23}} \nonumber \\
&&+\int d^2x\left[ V_2\Phi^2(x)+\alpha \rho (x)+\beta \rho (x)^2\right] ,
\end{eqnarray}
where $-f_0$ is the free energy of the system without spinon pairing. 

In the dual language, a phase incoherent state corresponds to a superfluid
state of vortices with  $<\Phi >=\Phi_0\neq 0$\cite{bfn}; 
while a long range phase
correlated state corresponds to a normal state of vortices with $<\Phi >=0$.
Thus to investigate the property of a quantum disordered spin gap phase,
only the superfluid phase of vortices needs be considered. In this case the
vortex field has a finite energy in its low-lying excitations, we can
therefore take a saddle point approximation for the $\Phi $ field. The
saddle point $\Phi_0$ is determined by the equation ${\delta S\left[ \Phi
(x),\rho (x)\right] /\delta \Phi (x)}=0$. In the small $\Phi (x)$ limit
and using the slow varying condition of both $\kappa_0(x)$ and
$\Phi(x)$, the equation becomes
\begin{equation}
-{\frac 43}f_0\left[ {\frac 12}\chi_d\kappa_0(x)\right]^{\frac 23}\Phi
_0^{\frac 13}(x)+2V_2(x)\Phi_0(x)=0,
\end{equation}

%%%%%%%%%%%%%%%%%%%%%%%%%%%%%%%%%%%%%%%%%%%%%%%%%%%%%%%%%%%%%%%%%%%%%%%%%
%%%The next paragraph is added by DX, Mar. 9, 1999.

Due to the presence of the first term, $\Phi(x)=0$ can not be 
the stable solution, 
which indecate that for any given configration of $\rho(x)$ the
effect of U(1) gauge field fluctuation will cause the condensation of 
vortice. 
In the description before duality, this means the quantum disorder
phase is always more stable than the ordered phase. The spinons become
"superconducting" when $\chi_d' \rightarrow\infty$, then in our approach
the pseudo gap phase is the middle state between the strange metal phase
in which $\chi_d'$ keeps constant with the decrement of temperature
 and the mean field pairing phase with
infinitive $\chi_d'$. At the same time, the phase transition at
$T^*_0$ predicted by the mean field theory becomes a crossover 
temperature $T^*$ below which the local minimum in the 
effective potential of $\rho$ moves away
from zero. 

Since the saddle point of $\Phi(x)$ is actually very large in small $\rho$
case, we must 
consider the saddle point equation in the large $\Phi (x)$
limit which leads to

\begin{equation}
\Phi_0^2(x)=\sqrt{\frac{4f_0}{{3\kappa_0(x)\chi_dV_2(x)}}}=4m^{*}c^2\sqrt{
\frac{f_0}{{3g^2\chi_d}}}\rho^{-2}(x).
\end{equation}

It can be shown that
this quantum disorder phase is more stable than the ordered phase. In the BFN's
theory, the condensation of vortices is obtained by assuming $V_2<0$. In our
case, however, $V_2$ is always positive and the condensation of vortices is
caused by the U(1) gauge field fluctuation. In the superconducting phase,
the U(1) gauge field is screened by the superfluid of holons, we find that $
\Phi_0(x)=0$. 

Taking the saddle point approxiamtion for the $\Phi $ field in $S\left[ \Phi
(x),\rho (x)\right] $, we then find the effective potential for the pairing
gap amplitude 
\begin{equation}
S\left[ \rho (x)\right] \approx -f_0+\int dx\left[ (\alpha +\alpha_0)\rho
+\beta \rho^2\right] . 
\end{equation}
where $\alpha_0=2\sqrt{g^2f_0/{3\chi_d}}$ is assumed to be weakly
temperature dependent. The $\alpha_0$ term is the contribution of the gauge
fluctuation. An important property revealed by this equation is that the
contribution from the gauge fluctuation to the free energy is proportional
to $\rho $, rather than $\rho^{5/6}$ as for the case in which the phase
fluctuation vanishes. This means that the gauge fluctuation is greatly suppressed by
the phase fluctuation. Since the contributions from both the gauge
fluctuation and the pairing condensation are now proportional to $\rho $,
the spin gap phase is therefore stable below a characterizing temperature
\begin{equation}
T^{*}=T_0^{*}-\frac{\alpha_0}a, 
\end{equation}
which is the solution of the equation $(\alpha +\alpha_0)|_{T=T^{*}}=0$.

Well below $T^{*}$, $\rho $ becomes finite and its fluctuation can be
omitted. In this case, the system we study is similar to the ````Nodal
Liquid Phase'' of BFN\cite{bfn}. 
The only difference is that in our model the local
pairs are not formed by electrons but by spinons. 

The phase fluctuation modifies the Landau diamagnetic susceptibility. Under
the saddle point approximation, the renormalized Landau diamagnetic
susceptibility is approximately given by
\begin{equation}
\chi_d^{\prime }\approx \chi_d+<\left( {{\frac 12}\kappa_0\Phi_0^2}
\right)^{-1}>=\chi_d(1+u<\rho >),
\end{equation}
where $u=\sqrt{{3g^2/\chi_df_0}}$ and $<\rho >=\int_0^\infty d\rho \rho
e^{{-(\alpha +\alpha_0)\rho -\beta \rho^2}\over {k_BT}}$

In the gauge theory, the electronic resistivity $R(T)$ is determined by the 
the transport scattering rate of holons $\tau_B^{-1}$. In the strange
metal phase above $T^{*}$, $\tau_B^{-1}$ is determined by the Landau
diamagnetic susceptibility $\chi_d:$\cite{LN}
\begin{equation}
\tau_B^{-1}\approx \frac{k_BT}{m_B\chi_d}.
\end{equation}
Since $\chi_d=\chi_F+$ $\chi_b$ is mainly determined by $\chi_F$ which
is nearly temperature independent,  $R(T)$ thus depends linearly on $T$ in
this phase. In the spin gap phase, $\chi_d$ in $\tau_B^{-1}$ should be
replaced by $\chi_d^{\prime }$. Since $<\rho >,$ and subsequently $\chi
_d^{\prime }$, increases rapidly with decreasing temperature in the spin gap
phase, the temperature dependence of $\ \tau_B^{-1}$ is therefore changed.
Near $T^{*}$, the deviation of$\ R(T)$ from its high temperature linear
dependence is approximately given by 
\begin{equation}
\frac{R(T)}{CT}=\frac{\chi_d}{\chi_d^{\prime }}\approx 1 -u<\rho >,  
\end{equation}
where $C$ is the slope of $R(T)$ at high temperatures. From the previous
result of $<\rho >$, we find that the leading temperature dependence of $
\frac{R(T)}{CT}\ $ calculated by us fits well with the experimental data
of $YBa_2Cu_3O_{7-x}$\cite{res2} as shown in Figure 1.

%%%%%%%%%%%%%%%%%%%%%%%%%%%%%%%%%%%%%%%%%%%%%%%%%%%%%%%%%%%%%%%%%
%Added by DX Mar.9, 1999
In this paper, we proposed that due to the strong fluctuation
of the U(1) gauge field, the proper description of the pseudo gap phase
is the quantum disordered spinon pairing state without the Bose-condensation
of holons. The mean field spinon pairing state 
is proved to be instable when the U(1) gauge field fluctuation is
included\cite{UL2}. 
While in our new description beyond mean field, the
local U(1) symmetry which is broken in the mean field pairing state restores
due to the phase disorder in the spinon pairing order parameter. Then the 
phase transition in the mean field description become a cross over from weak
pairing fluctuation regime ($T>T^*$) to the strong fluctuation regime
($T<T^*$) in the present paper. We further devide the pairing fluctuation
to amplitude and phase part and treat them seperatively. By integrating out
the phase fluctuation by duality transformation, we obtain the effective
potential for the amplitude. The crossover temperature can be determined by
the minimum of the effective potential moving away from zero. For the 
temperature much lower than $T^*$, the minimum in the effective potential
is far away from zero and the fluctuation of the amplitude is no longer
be important. Then our result is quite similar with BFN\cite{bfn} 
except that
in our case the local pairs are formed by spinons and the U(1) gauge field
fluctuation plays a very important role to obtain the quantum disordered
phase. For temperature near $T^*$, we can calculate the slope of the
resistivity by considering the effect of amplitude
fluctuation and our result fits very well with the experimental data.  

Very recently Y. B. Kim and Z. Q. Wang\cite{zq}
 proposed that the mean field spin gap
phase can be stabilized by strong critical fluctuation of holons. Compared 
with their approach, ours works better in quite high temperature near $T^*$,
where the diamagnetic susceptibility of holons $\chi_b$ can be viewed as 
constant.

\end{multicols}

\begin{figure}[htb]
\begin{center}
%\framebox[55mm]{\rule[-21mm]{0mm}{43mm}}
\psfig{file=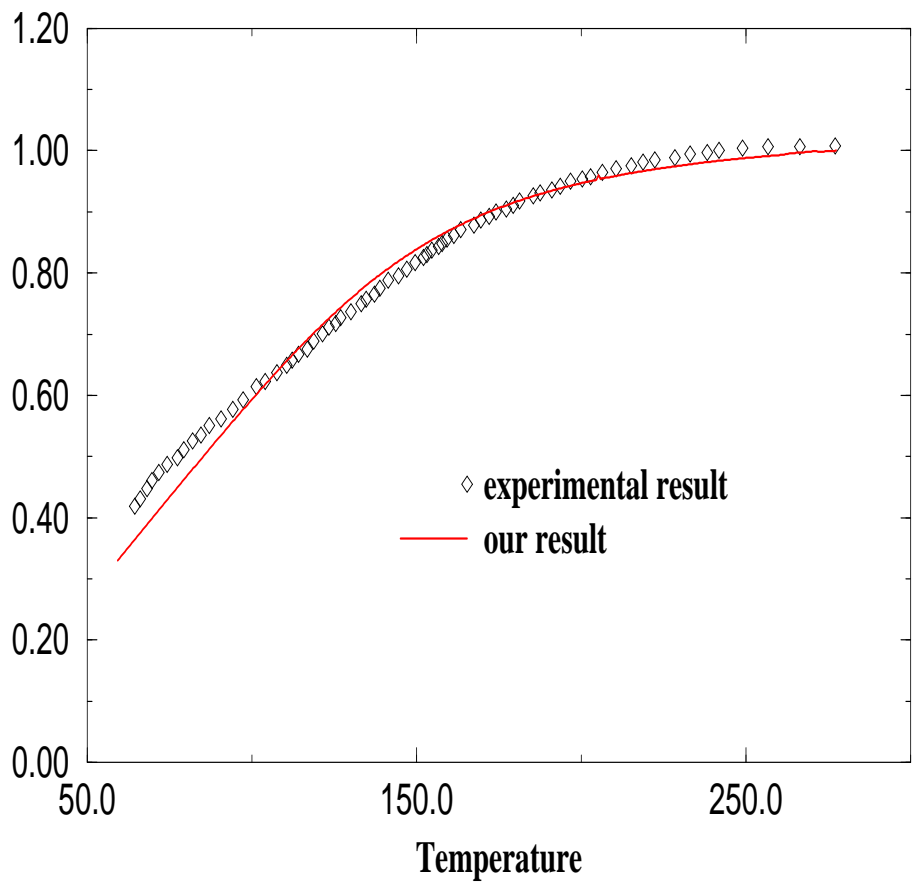,height=120mm,width=120mm,angle=0}
\end{center}
\caption{Comparison between our theoretical 
results of $\frac{\rho(T)}{{CT}}$ 
and the experimental data 
(square). The parameters used in fitting the experimental data
are $T^{*}=168K$, $a=19.1$, $\beta =975K$, $u=0.67$. }
\label{Fig.2}
\end{figure}

\end{document}